\documentclass{nature} 
\usepackage{graphicx}

\makeatletter
\let\saved@includegraphics\includegraphics
\AtBeginDocument{\let\includegraphics\saved@includegraphics}
\renewenvironment*{figure}{\@float{figure}}{\end@float}
\makeatother

\usepackage{float}
\usepackage{multirow}
\usepackage{subfigure}
\usepackage{mathtools}
\usepackage[colorlinks=true,linkcolor=blue,citecolor=black,urlcolor=blue]{hyperref}
\usepackage{authblk}
\usepackage[labelfont=bf]{caption}
\usepackage[figurename=Figure]{caption}
\usepackage{siunitx}  
\usepackage{soul} 
\usepackage{amsmath}
\usepackage{lineno}
\usepackage{tikz}
\usetikzlibrary{calc}
\usetikzlibrary{matrix}

\setstcolor{red}

\title{Unlocking Hidden Potential in Electron Holography of Non-Collinear Spin Textures}

\author{Moritz Winterott$^{1,2,*}$, Samir Lounis$^{1,2,3,*}$}

\begin{document}

\maketitle

\begin{affiliations}
    \item Peter Grünberg Institut, Forschungszentrum Jülich and JARA, D-52425 Jülich, Germany
    \item Faculty of Physics, University of Duisburg-Essen and CENIDE, D-47053 Duisburg, Germany
    \item Institut für Physik, Martin-Luther Universität Halle-Wittenberg, D-06099 Halle, Germany
    \item[*] Corresponding authors: m.winterott@fz-juelich.de, s.lounis@fz-juelich.de
\end{affiliations}

\section*{Abstract}
Due to their particle-like properties, three-dimensional (3D) spin textures have garnered significant interest, particularly for their potential applications in next-generation information storage devices. However, efficiently identifying these textures remains a major challenge. Here, we approach this problem from a new perspective. Rather than relying solely on the magnetic stray field, which vanishes in antiferromagnets, we use multiple-scattering theory to demonstrate that spin textures carry nontrivial charges due to the noncollinearity of magnetic moments. This induced charge encodes magnetic information driven by spin-mixing and spin-orbit interactions. We propose leveraging electron holography to extract this information by reconstructing phase images obtained from transmission electron microscopy (TEM). To quantify this effect, we systematically calculate and compare the contributions of both conventional and newly identified mechanisms to the phase images, considering different electronic structure parameters. Our findings mark a significant milestone in advancing the exploration and possible application of 3D spin textures in next-generation spintronic devices.

\section*{Introduction}

The miniaturisation of data storage units leads to increasing impact of quantum effects and thermal effects which hinders a further decreasing in size. Therefore the ongoing search for radically new storage techniques is vital to keep up with the industrial and academic needs of future devices. Particular interesting for future spintronic devices are topological protected spin textures, which are spatially localised quasiparticles.
Prominent examples are (anti-) skyrmions, chiral bobbers and hopfions.
The skyrmion is a topological spin-texture, with a whirlpool-like magnetic profile, confined in two dimensions \cite{article_Bogdanov1989,article_Bogdanov2006,article_Pappas2009,article_Neubauer2009,article_Muehlbauer2009}. Skyrmions can acquire a third dimension once stacked upon each other to form skyrmion-tubes~\cite{article_Milde2013,article_Birch2020,article_Seki2022}.
Compared to skyrmions, bobbers are relatively new spin textures~\cite{article_Rybakov2015,article_Zheng2018,article_Ran2021}, which consists of skyrmion-tubes, 
with decreasing size along one direction, ultimately terminating  
in a magnetic singularity, called a Bloch-point. 
Recently, another type of three-dimensional (3D) topological spin textures, the hopfions \cite{article_Whitehead1947,article_Tai2018,article_Sutcliffe2018}, became heavily explored due to their intriguing properties. The hopfion has a donut-like shape and can be seen as a twisted skyrmion-tube, where the two ends are connected to form a closed loop. It was predicted to be hosted in frustrated \newline
magnets~\cite{article_Sutcliffe2007}, and has been observed recently~\cite{article_Kent2021,article_Kiselev2023}. 
We note however that the 
identification and visualization of 3D spin textures, especially for hopfions, remain challenging.

Powerful experimental tools utilized to explore such spin textures are X-ray techniques (X-PEEM, MTXM, etc.)\cite{article_Tonner1988,article_Stoehr1988} or transmission electron microscopy (Lorentz imaging \cite{article_Blackburn1969,article_Chapman1984}, electron holography\cite{article_Gabor1949,article_Tonomura1992,article_McCartney2007,
article_Midgley2009,inbook_Kasama2011}, etc.)   due to their significant spatial resolution. However, they only provide access to a portion of the projected spin-texture. 
In principle, a reconstruction of the 3D spin-texture is possible by obtaining images from different incident beam angles with electron-holography\cite{article_Wolf2019} and MTXM\cite{article_Rodriguez2020}. Naturally, the more information is obtained from each individual images, the more meaningful the information about the 3D structure becomes. 

\begin{figure}
    \centering
    \includegraphics[width=.99\textwidth]{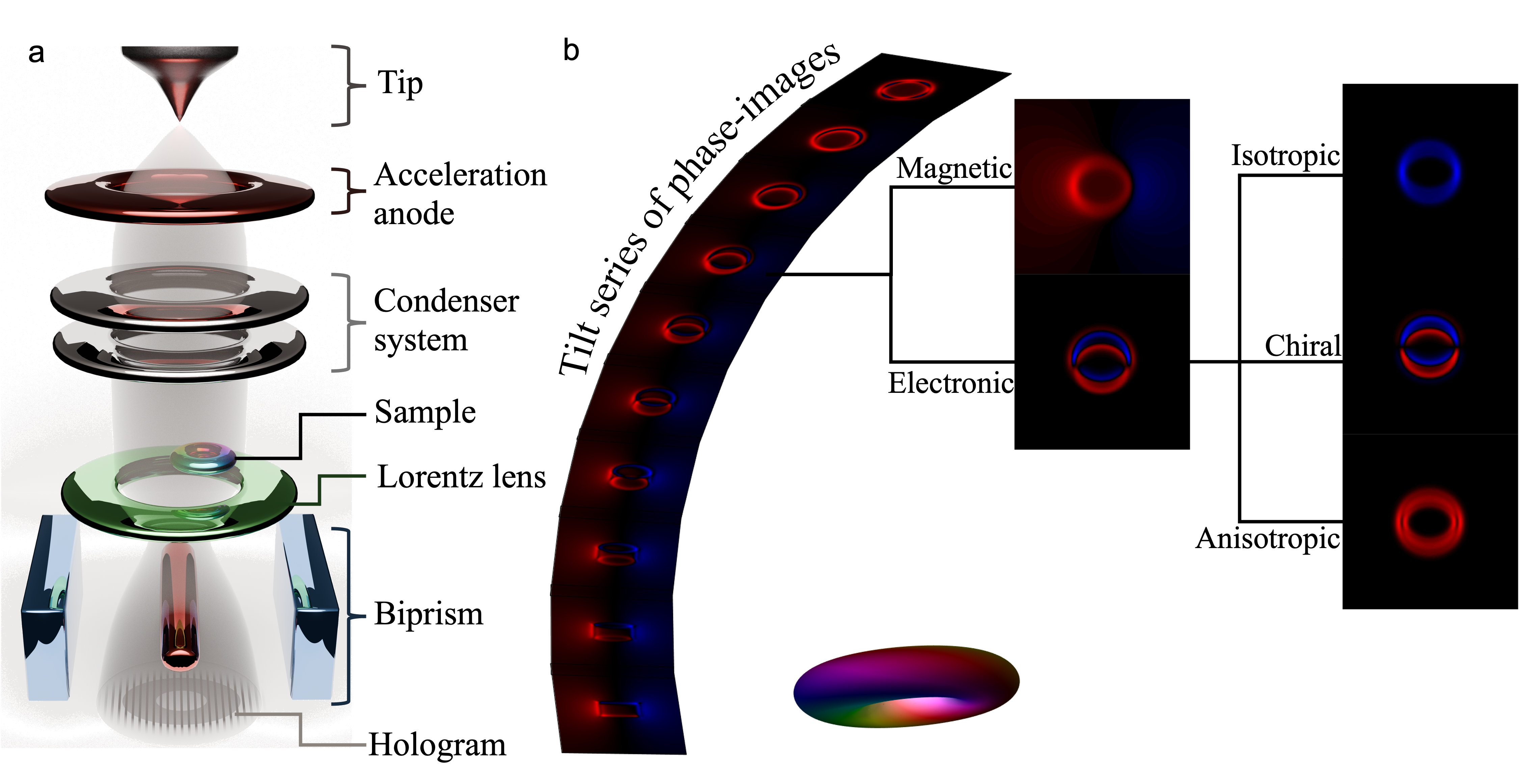}
    \caption{\textbf{Electron holography for complex spin textures.} (a) Schematic and simplified setup for capturing phase images within a transmission electron microscope. (b) Tilt series of phase-images and decomposition of a single phase-image into magnetic and electronic components as exemplified for a hopfion. The electronic component consist of three contributions that depend on the rotation of the magnetic moment and presence of spin-orbit coupling. For the simulation we used a common value of $300$ keV for the acceleration voltage.
    }
    \label{fig: panel1}
\end{figure}

In the current work, we demonstrate that electron holography images contain additional and nontrivial information, so far not considered, which are paramount for the characterization of complex spin textures (see Fig.~\ref{fig: panel1}). Up to now, the data is analysed in terms of the stray magnetization emanating from the explored spin textures. Here, we show, however, the emergence of a series of induced charges carried by the spin textures, which deflect the impinging electron-beam and contribute to the phase shift detected in transmission electron microscopy (TEM). The identified induced charges are shaped by isotropic, chiral non-collinear and spin-orbit contributions while remaining finite in antiferromagnetic (AFM) non-collinear materials. We discuss these new terms, their amplitude and impact on TEM images, which offer promising new opportunities in exploring 3D complex magnetic objects.

\section*{Results}

\subsection{Theory of magnetic induced charges.}

Here, we introduce the mechanisms leading to the emergence of the magnetic induced charges. Utilizing multiple scattering theory expressed in a tight-binding formulation, we quantify how the charge carried by a given magnetic atom is modified by the misalignment of the surrounding magnetic moments, as it is the case for a non-collinear spin-texture. 

We present one of the main results of this paper, that is the formulation of the non-collinearity induced charge as function of the angle between neighboring magnetic moments, systematically expressed in terms of dot products of magnetic moments (isotropic), cross products (chiral and linear with spin-orbit coupling) as well as mixed products involving the magnetic moments and the anisotropy field induced by spin-orbit coupling (SOC).

Assuming the following Hamiltonian describing without loss of generality, that each site hosts a single orbital as expressed in the Anderson model \cite{article_Anderson1961} amended with hopping \cite{article_Bychkov1984,article_Chaudhary2018} between nearest neighboring atomic sites: 
\begin{equation}
    \mathbf{H} = \sum_{j} \left[\epsilon_{d}- i \Gamma - U\mathbf{m}_{j} \cdot \mathbf{\sigma}\right]c^{\dagger}_{j}c_{j} + t \sum_{<i,j>}  \left[\cos{\varphi_{R}}-i\ \sin{\varphi_{R}}\ \mathbf{n}_{ij} \cdot \sigma \right] c^{\dagger}_{j}c_{i} \;.
    \label{eq: grn orbital}
\end{equation}

In the first term of eq.~\ref{eq: grn orbital}, $\epsilon_{d}$ corresponds to the orbital's energy (in the context of our current work $d$-orbitals carry the magnetic moments) before the energy splitting is induced by magnetism. The splitting is accounted for by the term $ U\mathbf{m}_{j} = U m\ \mathbf{e}_{j} \cdot \mathbf{\sigma}$, where $Um$ represents the exchange splitting, $U$ being the intra-atomic exchange interaction, $m$ the length of the magnetic moment, $\mathbf{e}_{j}$ its direction $\mathbf{\sigma}$ the vector of Pauli matrices and $c_{j}\, (c^{\dagger}_{j})$ stands for the annihilation (creation) operator for site $j$. While the magnetic moments can be misaligned, their magnitude, for simplicity, is assumed to be same for all sites.  $\Gamma$ is shaped by the coupling of the orbitals to the bath of electrons, which triggers a broadening of the associated electronic states. 

The second term of eq.~\ref{eq: grn orbital}  accounts for SOC and leads  to the Rashba Hamiltonian~\cite{article_Bychkov1984,article_Chaudhary2018}, which comprises both a non-magnetic and chiral Rashba components. The connection to the Rashba model is detailed in Supplementary Note 1. 
The Rashba-angle $\varphi_{R}$ fine-tunes the balance between the conventional spin-independent and chiral Rashba hopping. 
The anisotropy field $\mathbf{n}_{ij}$ varies by site and emerges from SOC, leading to a broken inversion symmetry since $\mathbf{n}_{ij} = -\mathbf{n}_{ji}$. It defines a chiral vector, similar to the Dzyaloshinskii Moriya interaction (DMI)\cite{article_Dzyaloshinskii1957,article_Moriya1960,article_Moriya1960_2}.

\begin{figure}[h]
\centering

\begin{tikzpicture}
    \matrix (contribution) [matrix of nodes,nodes={minimum width=1cm,minimum height=1cm},inner sep=3mm]
    {
    & \node{isotropic}; & \node{chiral};      & \node{anisotropic};    \\
  \node {}; & $(\cos^2{\varphi_{R}}-\sin^2{\varphi_{R}})(\mathbf{e}_{i} \cdot \mathbf{e}_{j})$ & $\cos{\varphi_{R}}\sin{\varphi_{R}}(\mathbf{e}_{i} \times \mathbf{e}_{j})\cdot \mathbf{n}_{ji}$ & $ \sin^2{\varphi_{R}} (\mathbf{e}_{j}\cdot \mathbf{n}_{ji}) ( \mathbf{e}_{i} \cdot \mathbf{n}_{ji})$ \\
  \node {}; & \includegraphics[width=.1\textwidth]{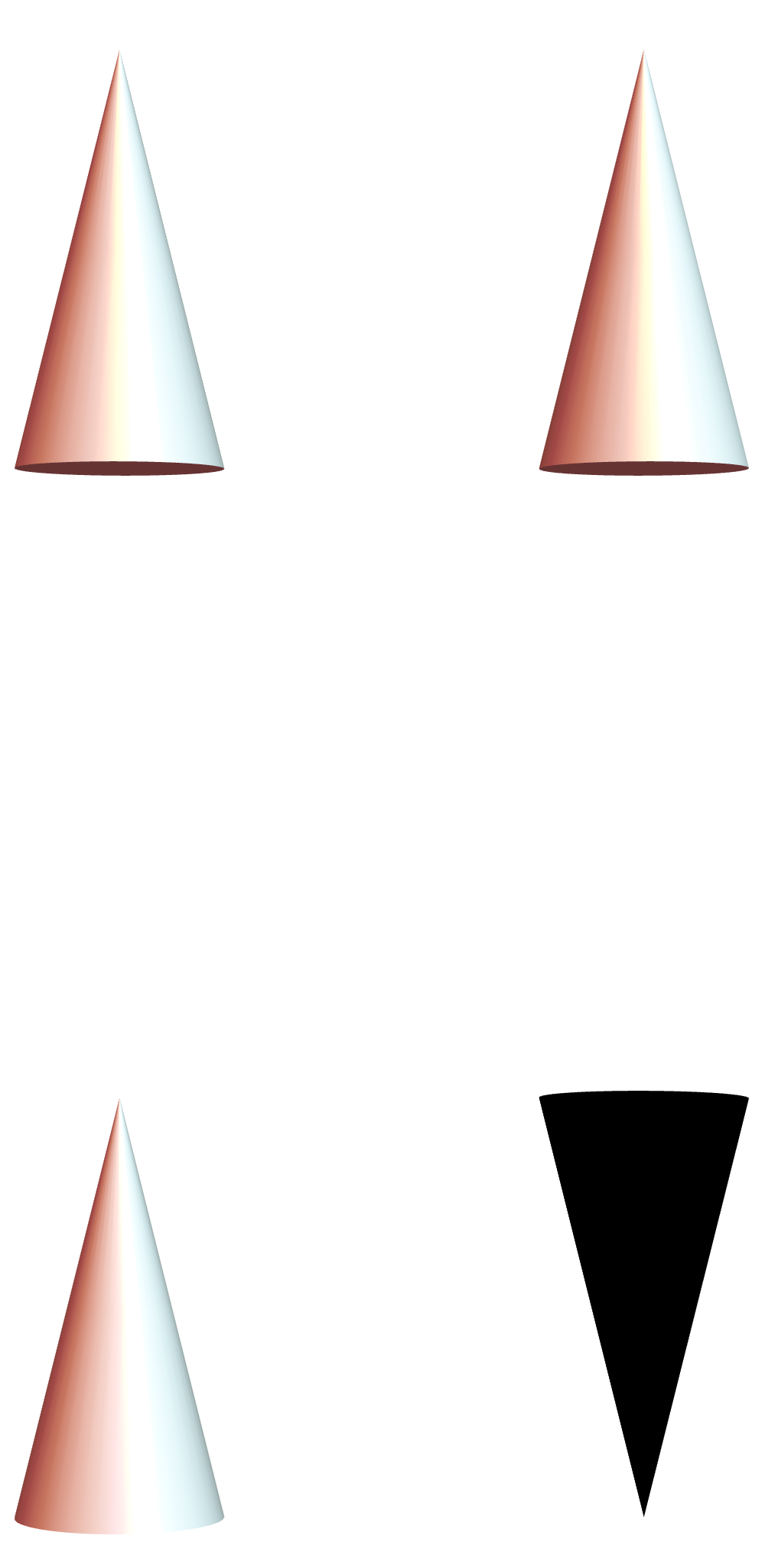}                          & \includegraphics[width=.1\textwidth]{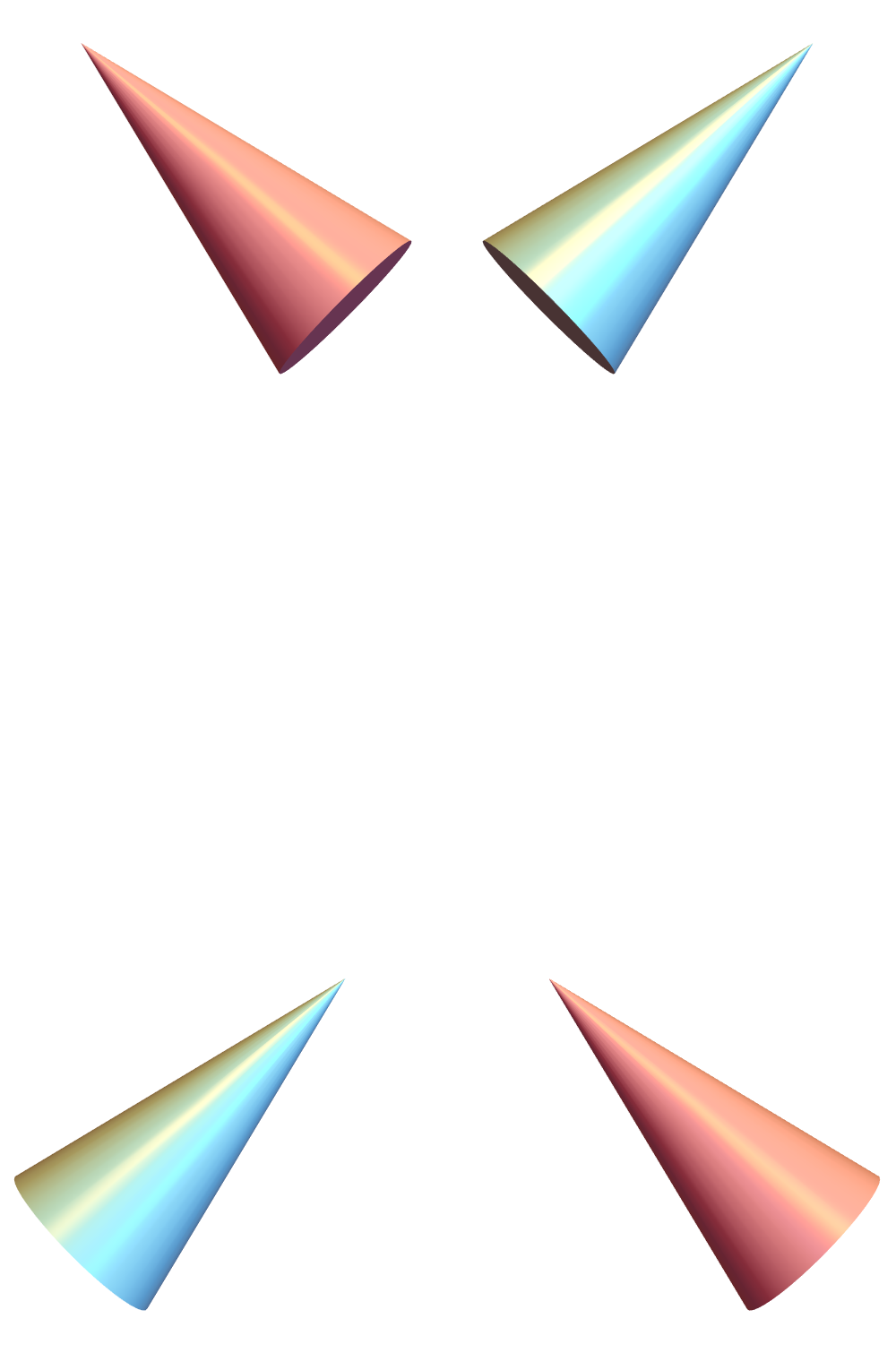} 
            & \includegraphics[width=.1\textwidth]{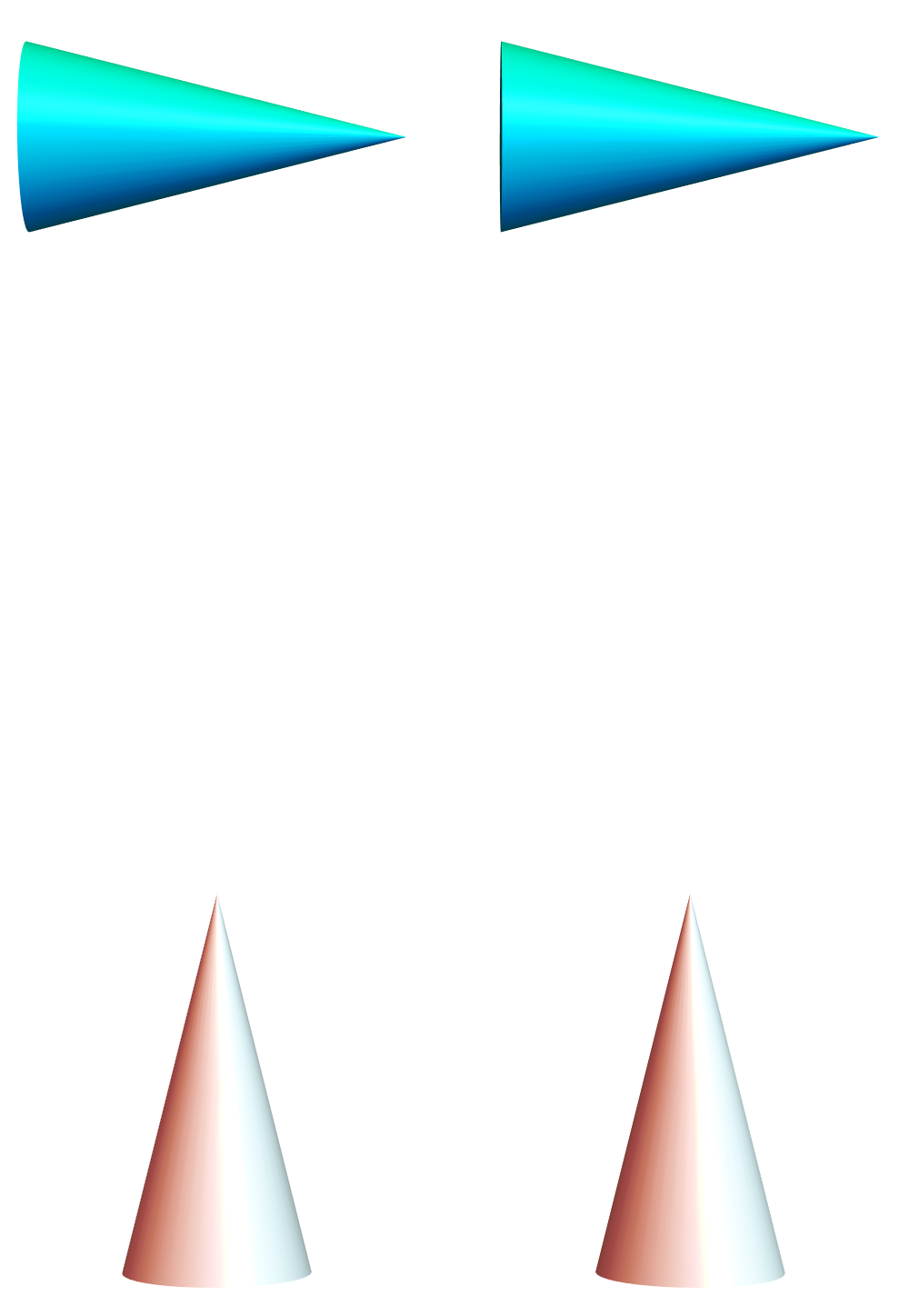}\\
  };
    
    \node[anchor=south, xshift = 5mm, yshift = -5 mm, rotate=90] at (contribution.west) {};
    \node[anchor=south, xshift = 6mm, yshift = -6 mm] at (contribution.north) {};
\end{tikzpicture}

\caption{\textbf{Tabulation showcasing the isotropic, chiral and anisotropic contributions to the non-collinearity induced charge.}
The arrows indicate the type of magnetic states that can be distinguished by the different terms. For example, the isotropic induced charge differentiates the ferromagnetic and antiferromagnetic configurations. The chiral term monitors the change of chirality while the anisotropic one senses the rotation of both moments with respect to the SOC anisotropy field.}
\label{tab: grn expansion}
\end{figure}
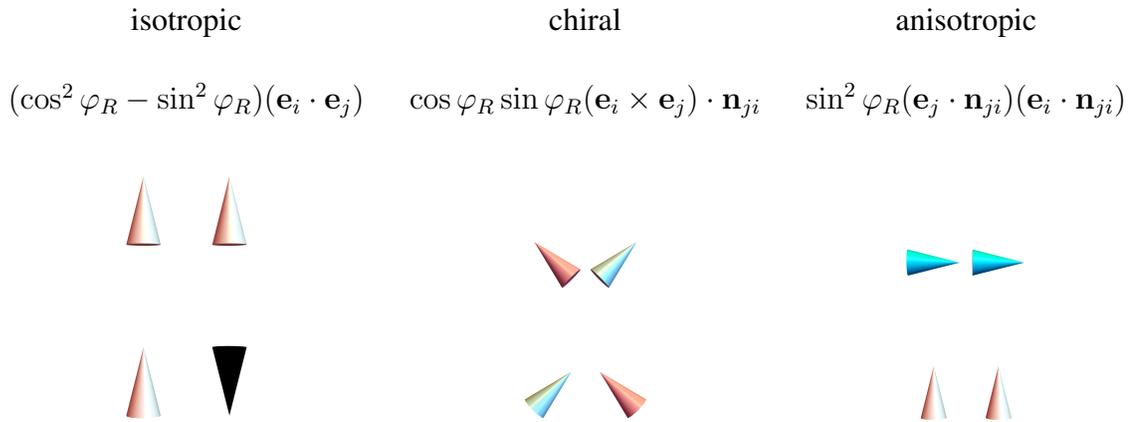

Based on a multiple-scattering approach, utilizing the hopping in a perturbative framework (Supplementary Note 2), we identify a series of charges induced at site $j$ by the misalignment of magnetic moments located at sites $i$. 
We focus on processes that are second-order in terms of the hopping $\mathbf{t}$, which are tabulated in Fig.~\ref{tab: grn expansion}.

The first contribution in Fig.~\ref{tab: grn expansion} is termed isotropic due to its cosine dependence on the angle between the two magnetic moments; the cosine function cannot differentiate the sign of the angle. It hinges on the difference between spin-independent and chiral Rashba hopping. While being finite without SOC, this term gets reduced with the possibility of cancellation or even sign reversal once incorporating the Rashba coupling.  Thus, it emerges as a hybrid of a 0th-order and 2nd-order contribution in SOC, necessitating two electron scattering events between two magnetic moments. Interestingly, this isotropic induced charge is finite for both cases, ferromagnetic and antiferromagnetic alignement of the moments. 

The second contribution is chiral since it depends on the cross-product of the two magnetic moments and is 1st-order in SOC. To be finite, it needs broken inversion symmetry. This allows the chiral contribution to be sensitive to the chiral nature of the underlying 
spin textures. The last contribution 
is anisotropic and  2nd-order in SOC while having a form similar to that of an anisotropy energy, but generalized to a non-local scattering, i.e. involving two different magnetic moments. 
 For large spin textures, and taking micromagnetic limit, one can show that the local contribution prevails (Supplementary Note 3). Furthermore, in Supplementary Fig. 2 we address the accuracy of the perturbative approach by making a comparison to the exact result for a dimer.

Armed with our findings, we can quantify the different contributions for arbitrary complex spin profiles. In the next subsection, we establish a link to experiments by addressing the impact of non-collinearity induced charges to TEM and electron holography.

\subsection{Scheme to quantify phase images and electron holography.}
\label{subsec: EH}

In the preceding subsection, we observed that a spin-texture leads to the redistribution of charge. This, as we show in the following subsections, can be used to distinguish a spin-texture from the ferromagnetic or antiferromagnetic surrounding. 
In this section, we present a scheme that enables to evaluate the phase images induced by the newly induced
charges and captured by electron holography within TEM.. The latter offers an ideal playground for detecting such
induced charges, owing to its remarkable 
spatial resolution, sensitivity to electrostatic potential,
and its existing use for the identification of 3D spin textures when probing the impact of stray magnetic  field.\cite{article_Zheng2025}.

This technique involves directing an electron wave onto the sample. As it traverses the material, the wave accumulates a phase shift, denoted as 
$\varphi$, due to scattering at the material's electrostatic and magnetic vector potentials. While conventional transmission microscopy focuses on measuring the electron wave's amplitude, thereby losing phase information, electron holography allows to recover the phase data. 
There exists several techniques for electron holography, here we address the common off-axis electron holography \cite{article_Midgley2001}. Thereby two-dimensional phase image are obtained which corresponds to both the electrostatic potential $\Phi$ and the magnetic vector potential $\mathbf{A}$:
\begin{equation}
\varphi(x,y) = C_{E} \int \Phi(x,y,z)  \text{d}z - \frac{e_{0}}{\hbar} \int \mathbf{A}_{z}(x,y,z)  \text{d}z,
\label{eq: eh phaseshift definition}
\end{equation}
where we have assumed for simplicity that the electron wave propagates along the $z$-direction. $e_{0}$ represents the absolute value of the electron charge. The prefactor $C_{E}$ for the electronic contribution  depends on the acceleration voltage used in the TEM, which 
for the purpose of our study is a major advantage over the magnetic component, as it allows the relative strength of the electronic component to be tuned. 

As aforementioned, so far, the electronic contribution is considered to be transparent to the underlying magnetic textures when probed by electron holography, which is contradicted by our derivations from the precedent subsections. In Fig.~\ref{fig: panel1} b the schematic construction of phase images is shown. A spin-texture creates a phase-images which depends on the angle of the incoming electron wave. For each angle, the phase-images consist of several contribution which are either magnetic or electronic. We focus on the electronic contributions, which arises from the contributions shown in Fig. \ref{tab: grn expansion}. It is at this stage important to be able to compare the relative strength between the newly proposed electronic contribution to the conventional magnetic one. The latter can be evaluated by an optimized forward model \cite{thesis_caron2017} that has already demonstrated success \cite{article_Denneulin2021}, and  reproduced in the following.

Starting from equation \ref{eq: eh phaseshift definition} and assuming for simplicity that the electron waves travel along the $z$-direction, one can rewrite the magnetic contribution as
\begin{align}
    \varphi_{mag}(x,y) &= \frac{-\mu_{0} M_{sat}e_{0}}{\hbar} \int \int \frac{(y-y')m_{pr,x}(x',y')-(x-x')m_{pr,y}(x',y')}{(x-x')^{2}+(y-y')^{2}} \text{d}x' \text{d}y'\, , \\
    \textbf{m}_{pr}(x,y) &= \int \mathbf{m}(x,y,z) \text{d}z \, ,
    \label{eq: eh mag phaseshift1}
\end{align}
where $M_{sat}$ is a  material dependent saturation magnetization and $\mu_{0}$ the vacuum permeability. Such an equation can be readily solved numerically using e.g. the convolution theorem.

The electronic contribution demands the development of a new method, one that is consistent with the approach used for the magnetic part (see Supplementary Note 4 for details). 
We start from eq. \ref{eq: eh phaseshift definition} and insert the Poisson’s equation:
\begin{align}
  \varphi_{elec}(x,y) &= \frac{C_{E}}{ 4 \pi \epsilon_{0} \epsilon_{r}} \int \rho(\mathbf{r}') \int \frac{e^{-\lvert\mathbf{r}-\mathbf{r}'\rvert k_{tf}}}{\lvert\mathbf{r}-\mathbf{r}'\rvert} \text{d}z \, \text{d}^3r' \,,
  \label{eq: yukawa}
\end{align} 
which incorporates screening effects via the Thomas-Fermi screening parameter $k_{tf}$.
We are now equipped to explore  spin textures addressed in the next subsection.

\subsection{Non-collinarity induced charges of 3D spin textures.}

We explore in this section the charge induced by  diverse 3D spin textures. We start with the skyrmion-tubes, since they have been heavily investigated experimentally and are therefore the ideal candidates to test our predictions. 

\begin{figure}[H]
    \centering
    \includegraphics[width=.99\textwidth]{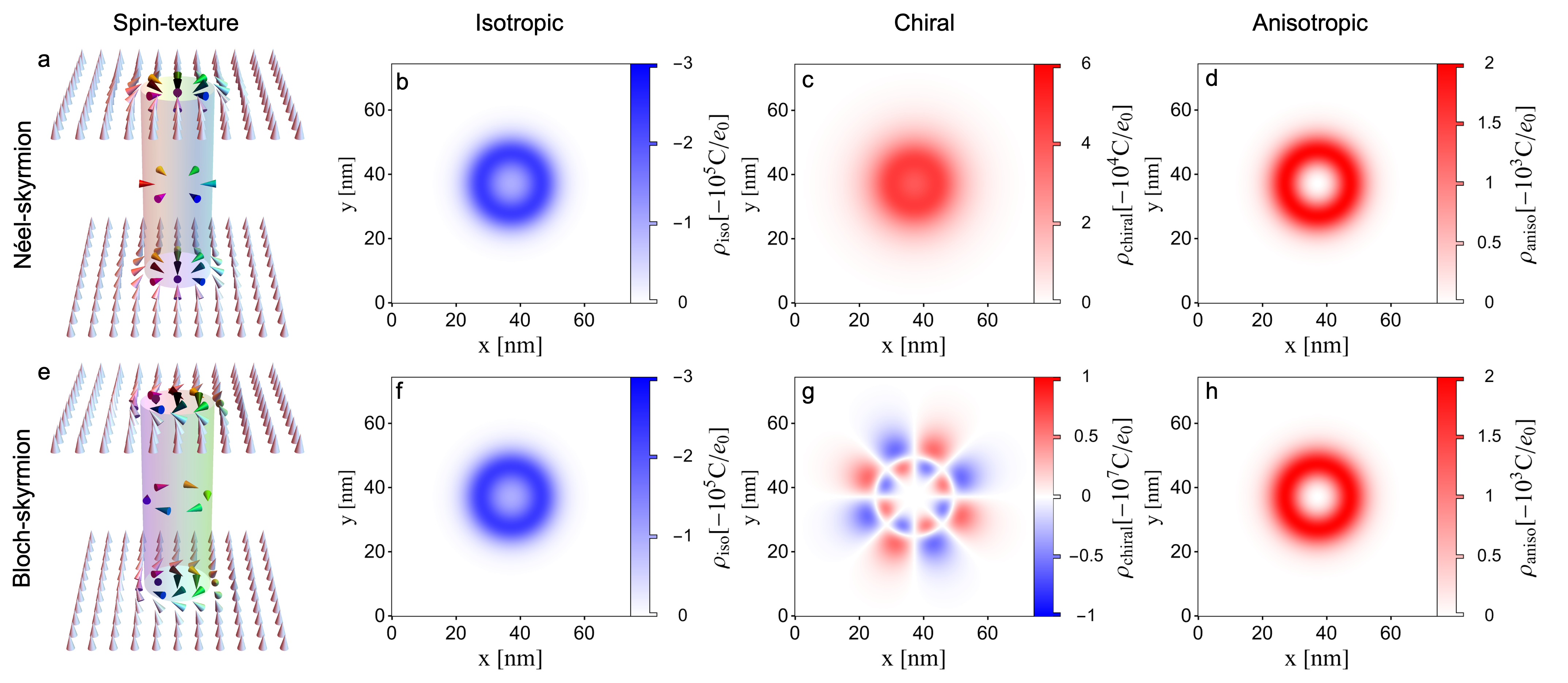}
    \caption{\textbf{Non-collinearity induced charges and spin-profiles for a Néel-type and Bloch-type skyrmion-tube.} The top row is associated to the Néel-type skyrmion-tube, while the bottom row shows the Bloch-type skyrmion-tube. The patterns of the induced charges is rich and can dramatically change depending on the type of skyrmions. The skyrmion profile is obtained \newline
    from \cite{article_Roland2015} with $c=w= 10$ nm.
    }
    \label{fig: panel2}
\end{figure}

\begin{figure}[H]
    \centering
    \includegraphics[width=.99\textwidth]{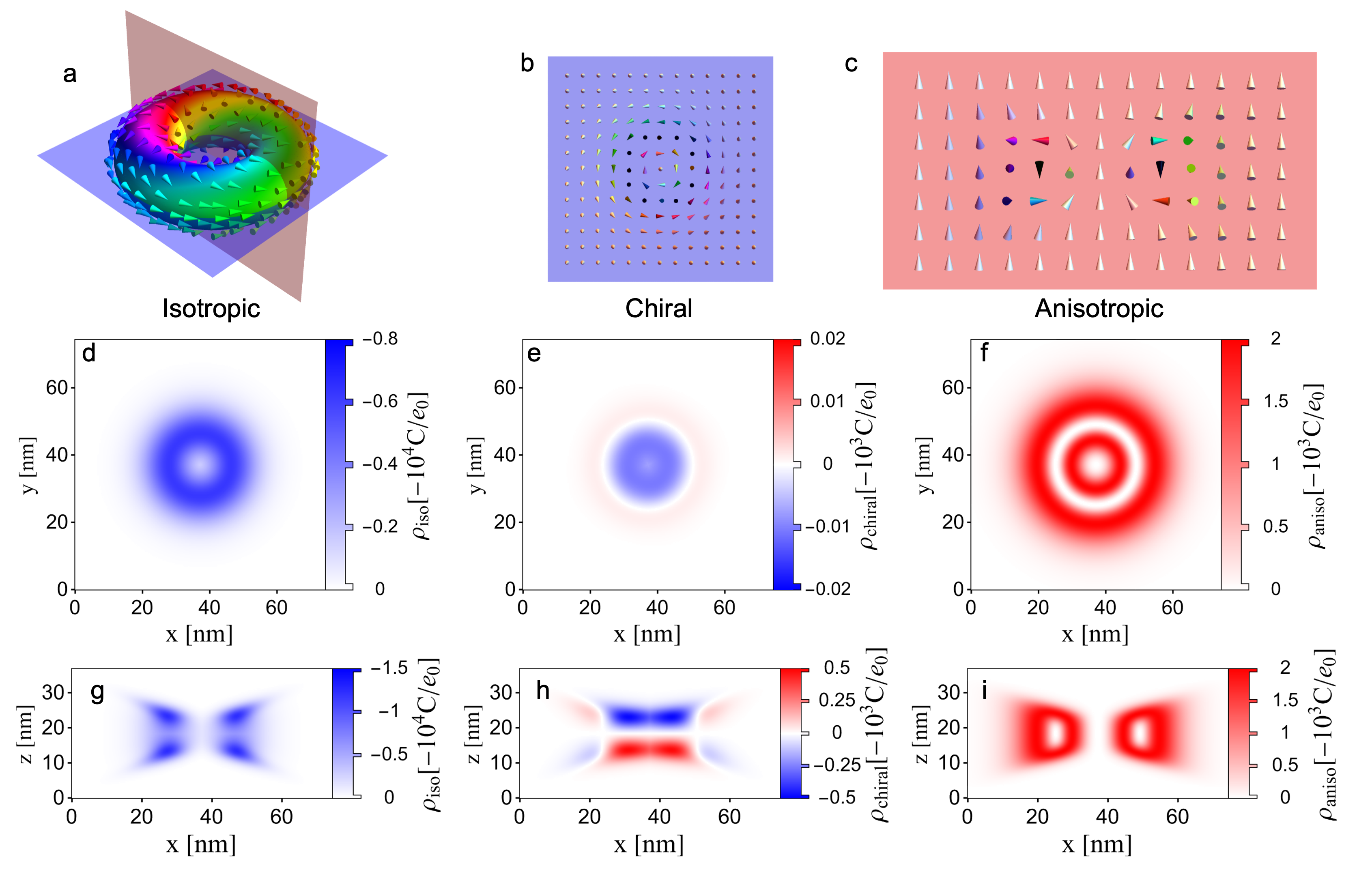}
    \caption{\textbf{Visualization of a hopfion with the associated non-collinearity induced charges.} (a) the spin-profile on the surface of a torus (hopfion  with hopfindex $1$) is shown. Two planes are depicted which correspond to the cuts shown in (b) and (c).  (d, e, f) show the three contributions to the induced charge for cut shown in (b), while (g, h, i) correspond to cut (c). The hopfion profile is obtained from \cite{article_Wang2019} which is based on the ansatz presented in \cite{article_Hietarinta1999}. Thereby we use $R= 12\,\text{nm}, w_{R}=7.5\,\text{nm}, h=6\,\text{nm and } w_{h}=3.25\,\text{nm}$.}
    \label{fig: panel3}
\end{figure}

In Fig.~\ref{fig: panel2}, we plot the different non-collinearity induced charges for the Néel- and Bloch-type of skyrmion tubes. As an example, we assume the following electronic parameters: $Um = 0.677$ eV, $\epsilon_{d} = 0$ eV, $t = 0.07$ eV, $\Gamma = 0.15$ eV,  $\varphi_{R} = \pi/12$ and the Fermi energy $\epsilon_{F}$ is set to $0.48$ eV.
The isotropic contribution (b and f) forms a ring pattern which is identical for both types of skyrmions. This is due to the underlying mechanism giving rise to the isotropic induced charge, namely the cosine angle between neighbouring sites is identical for the two skyrmions. At the core of the skyrmions, the patterns is significantly weak, due to the underlying magnetic collinearity. At the skyrmion edges, where the non-collinearity is maximized, the induced charge reach significant values. The chiral nature of the two magnetic objects being distinct, the chiral induced charge patterns are strikingly different (c and g). This is due to the anisotropy field giving rise to the chiral vector defining the chiral induced charge, which is within the Rashba model identical to the one giving rise to the interface-DMI.  
The anisotropic patterns, similarly to the isotropic ones, are identical for both skyrmion-tubes (d-h). 
After analysing the case of skyrmion-tubes, we address the case of hopfions. While skyrmion-tubes are characterized in our analysis by an identical spin textures along the $z$-direction, hopfions are different and therefore require us to consider the whole 3D charge distribution as shown in Fig.~\ref{fig: panel3}. As an example, we consider two cuts along the planes shown in Fig.~\ref{fig: panel3} a.

Starting with the isotropic patterns (d) (associated to in-plane cut (b)) in Fig.~\ref{fig: panel3}, we notice a isotropic disk located at the center of the hopfion. The perpendicular cut (g) reveals that the disk extends into a tube with increased intensity at the both ends. The chiral patterns posses a more complicated pattern. Two Regions with different sign emerge in the in-plane cut, however, the perpendicular cut reveals that these region changes sign along the z-axis, resulting in 4 regions in total. 
The anisotropic profile hosts ring-like features in (f) associated to the in-plane cut (c), with weak intensity in the center due to the alignment with be background in this region. 
Within the out-of-plane cut (c), the anisotropic shapes consist of two elongated features, with a mirror symmetry in the middle, which can be easily related to the spin textures depicted in (c). 
Indeed, in the region where the magnetic moments point along (white arrows) or against (black arrows) the anisotropic contribution is zero. 

After the analysis of the non-collinarity induced charges of various 3D spin textures, we address in the next subsection the associated phase images (see framework in section \ref{subsec: EH}) when probed with electron beams pertaining to TEM.

\subsection{Phase images and electron holography of 3D spin textures.}

\begin{figure}[H]
    \centering
    \includegraphics[width=.99\textwidth]{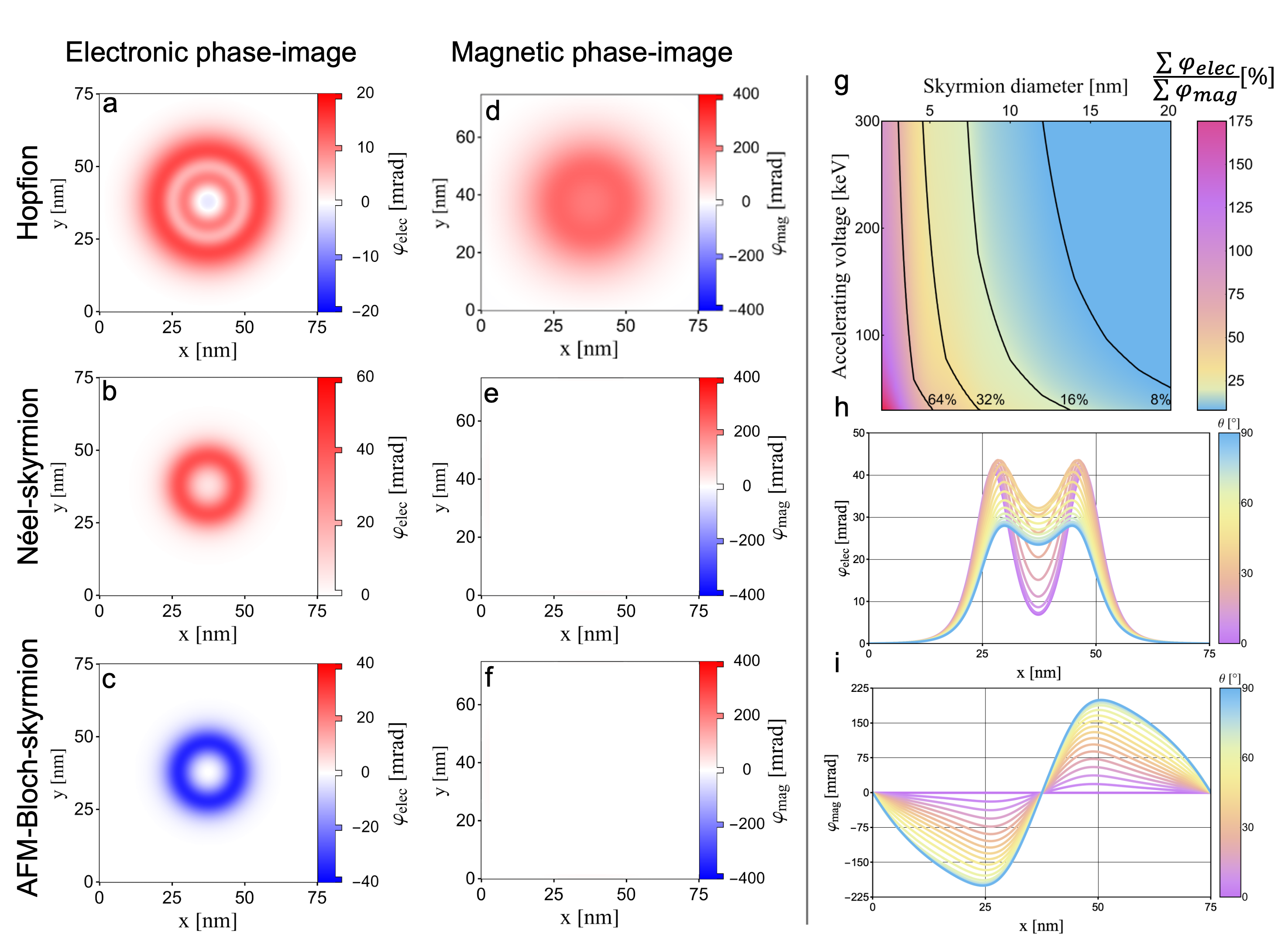}
    \caption{\textbf{Phase-images of different spin textures and impact of TEM acceleration voltage}. The left column (a-c) shows the electronic contribution to the phase-image from eq. \ref{eq: yukawa}, while the middle column shows the magnetic contribution to the phase-image (d-f), defined in eq. \ref{eq: eh mag phaseshift1}. The rightmost column pictures a comparison of the total electronic and magnetic contribution to the phase-image for a ferromagnetic Bloch-type skyrmion-tube with different skyrmion diameters and acceleration voltages used in the TEM (g). The line-profiles for the electronic (h) and conventional magnetic (i) phases along the center of a ferromagnetic Néel-type skyrmion-tube (d) are shown for different incident beam angles. As an example three different spin textures are used, a and b showcased phase-images for a hopfion, b and e for a Néel-type skyrmion-tube. The phase-images c and f are obtained for a antiferromagnetic Bloch-type skyrmion-tube.
    }
    \label{fig: panel4}
\end{figure}

The phase-images are tightly connected to the underlying non-collinearity induced charges.  This can readily be noticed when analysing the phase-image associated to the electronic contribution of the hopfion, see Fig.~\ref{fig: panel4} a, which is characterized by red  rings directly connected to the charge plotted in Fig.~\ref{fig: panel3} f. The latter panel showcases region with different sign, which are suppressed in the phase-image by the strong anisotropic contribution. Similar conclusions can be drawn for the the Néel-type skyrmion-tube when comparing the phase-image in Fig.~\ref{fig: panel4} b with the corresponding charge (fig. \ref{fig: panel2} g). For the acceleration voltage we use $300$ keV.

Looking at the magnetic contributions to the phase-images in Fig.~\ref{fig: panel4} (d-f), one sees that the Néel-type skyrmion-tube has a vanishing magnetic contribution for a electron beam perpendicular to the skyrmion. Experimentally, this was observed and therefore samples hosting such skyrmions are rotated to induce a non-vanishing magnetic signature \cite{article_Denneulin2021}.
Bloch-type skyrmion-tubes induce a magnetic signal in the phase-image without the need of rotation \cite{article_Park2014}. For a antiferromagnetic (AFM) Bloch-type Skyrmion-tube, however, atomic resolution is needed in order to pick up a finite magnetic signal (f) since the stray field cancels out.
The new presented electronic contribution to the phase-image does not suffer from such shortcoming and stay valid up to a sign change(c). This holds true for all con-collinear AFM spin textures. Comparing the impact of rotating the sample on the phase-images shown in Fig.~\ref{fig: panel4} (h-i),  one immediately notices striking differences. While the charge induced phases are rather finite and stable, the magnetic signal can even vanish.

Experimentally, we expect both signals, magnetic and charge-induced to be measured. To differentiate the two, 
 we explore how the acceleration voltage affects the magnitude of the charge-induced phases. As a reminder the phase induced by the magnetization stray field is voltage-independent. In Fig.~\ref{fig: panel4} (g), the ratio of of the total electronic and magnetic contribution to the phase-image are shown 
as function of the acceleration voltage $u$ and diameter $d$ of a spin-texture, here the Bloch-type skyrmion-tube. While typically an acceleration voltage of 300 keV is used, we predict that reducing both the voltage and skyrmion diameter enhance the amplitude of the charge-induced phase. Relatively low voltages (even below 100 keV) can induce significant charge-induced phases.  
This is a general behavior for the isotropic and chiral contribution, which holds also true for different spin textures.

\section*{Conclusion}

 By demonstrating that the charge pertaining to a magnetic state changes in a non-trivial fashion when the magnetic moments are perturbed, we identified analytically that the underlying induced charge follows specific dependencies with respect to the angle between the magnetic moments. This implies that the electronic signal, not considered so far when analysing experimentally TEM electron-holographic images, contains crucial information on non-trivial magnetic states. 
 
 Our findings unveilling isotropic, anisotropic and chiral  contributions to the images accessible experimentally open unprecedented perspectives in the realm of 3D spin textures probed via TEM. Similar concepts have been shown to be crucial for the vizualisation of spin textures via scanning tunneling spectroscopy~\cite{article_Imara2022,article_Crum2015,article_Hanneken2015}.
 
The textures of the phase images is rich and distinct when comparing the conventional magnetic ones to the charge-induced ones. The interplay of both contributions to the total signal can be complex depending on the underlying spin textures. Here, we addressed and compared signals pertaining to N\'eel and Bloch skyrmion-tubes as well as hopfions. An appealing procedure to identify the newly proposed signal is to proceed to a study as a function of the acceleration voltage of the TEM's electron beams and size of the spin textures. The latter can be controlled by applying a magnetic field \cite{article_Roland2015}. Furthermore, a simple way to remove the magnetic contribution consists in a full rotation of the sample followed by a subtraction of the total signals (before and after rotation).
 
 So far the charge-induced signals were evaluated without accounting for self-consistency in the response of the charge to the magnetic texture. This assumption is more justified for large spin textures with a relatively small misalignment of the magnetic moments which is than expected to be associated with a small change in charge per site. This, however, does not affect the generality of our findings, namely the different types of dependencies with respect to the angles between the magnetic moments. In Supplementary Note 2, we compare the pertubative approach with the exact solution for a dimer.

 The surge for exploring complex spin textures in 3D calls for the invention of new experimental methodologies. Here, we demonstrate that "conventional" TEM setups can readily be utilized to explore an intriguing emergent phenomenon, overlooked so far. We foresee new developments that will enable the observation of magnetic-induced charges, which will be one additional milestone for the identification of 3D spin textures and their potential use as magnetic bits in information technology.
 
\begin{addendum}

\item
We acknowledge fruitful discussions with Rafal Dunin-Borkowski. 
This work was supported by the SPP 2137 “Skyrmionics” (Project LO 1659/8-1, S.L.) of the Deutsche Forschungsgemeinschaft (DFG).

\item[Author contributions]
S.L. initiated, designed and supervised the project. M.W. derived the theoretical framework and performed the simulations. Both authors wrote the manuscript.

\item[Competing interests]
The authors declare no competing interests.

\item[Data availability] The data that support the findings of this study are available from the corresponding authors on request.

\item[Code availability] The  code that supports the findings of this study is available from the corresponding author on request.
\end{addendum}

\clearpage

\section*{References}
\bibliography{references}
\end{document}

\begin{center}
    \Large{\textbf{\underline{Supplementary Materials}} }
    
    \Large{\textbf{Unlocking hidden potential in electron holography of non-collinear spin textures}}
    
    \large{Moritz Winterott and Samir Lounis}
\end{center}

\newpage 

\section*{Supplementary Note 1 - Rashba model.}
Starting from eq.~1 of the main text, we focus on the hopping~\cite{article_Chaudhary2018} contribution to the Hamiltonian:
\begin{equation}
    \mathbf{t} =t \sum_{<i,j>}  \left[\cos{\varphi_{R}}-i\ \sin{\varphi_{R}}\ \mathbf{n}_{ij} \cdot \sigma \right] c^{\dagger}_{j}c_{i},
    \label{eq: hopping}
\end{equation}
and shortly demonstrate the connection to the Rashba model~\cite{article_Bychkov1984}.
 Assuming for simplicity a 2-dimensional simple cubic lattice and periodicity, we perform a Fourier transformation, we obtain
\begin{equation}
    \mathbf{t}= 2 t \sum_{\mathbf{k}}  \left[\cos{\varphi_{R}} (cos{k_{x}}+cos{k_{y}})+\ \sin{\varphi_{R}}\  (\sigma_{y} \sin{k_{x}}-\sigma_{x} \sin{k_{y}}) \right] c^{\dagger}_{\mathbf{k}}c_{\mathbf{k}}\;,
    \label{eq: hopping-k}
\end{equation}
where $\mathbf{k}$ is given in units of the inverse lattice constant. 

For small values of $\mathbf{k}$ we recover the Rashba-type formulation of the Hamiltonian
\begin{equation}
    \mathbf{t} \approx -t \sum_{\mathbf{k}}  \left[\cos{\varphi_{R}} (k^{2}_{x}+k^{2}_{y})- 2\ \sin{\varphi_{R}}\  (\sigma_{y} k_{x}-\sigma_{x} k_{y}) \right] c^{\dagger}_{\mathbf{k}}c_{\mathbf{k}} + \text{constant}\,,
    \label{eq: hopping-rashba}
\end{equation}
which can be compared to the usual form: $\sum_{\mathbf{k}}  \alpha_{R} (\sigma_{x} k_{y}-\sigma_{y} k_{x})\, c^{\dagger}_{\mathbf{k}}c_{\mathbf{k}}$ thereby identifying the so-called Rashba-parameter $\alpha_{R}$ to be proportional to the sine of the Rashba-angle.

\section*{Supplementary Note 2 - Perturbative framework.}
\label{supplementary_Dyson_equation}

We separate the Hamiltonian in two different contributions: local terms $\mathbf{H}$ and hopping terms $\mathbf{t}$, containing the interaction between different atomic sites. The Green function associated to the non-interacting system $\mathbf{g}$ is known or can be easily evaluated. It can be used to construct the interacting Green function $\mathbf{G}$ by solving the Dyson equation
 \begin{align}
    \label{eq: grn dyson closed}
    &\mathbf{G}^{-1} = \epsilon\mathbf{1}-\mathbf{H}-\mathbf{t} = \mathbf{g}^{-1}- \mathbf{t} \\
    &\Rightarrow \mathbf{g} = \mathbf{G} -\mathbf{g} \mathbf{t} \mathbf{G} \Leftrightarrow \mathbf{G} = \mathbf{g} + \mathbf{g}\mathbf{t} \mathbf{G} =\frac{\mathbf{1}}{\mathbf{1}- \mathbf{g} \mathbf{t}} \mathbf{g}\; ,
\end{align}
where we have used 
 \begin{align}
    \label{eq: grn hamilton operator components 2}
    \mathbf{g} = (\epsilon\mathbf{1}-\mathbf{H})^{-1}\;.
\end{align}
Obtaining the Green function clearly involves matrix inversion and multiplication which can be computationally extremely challenging for huge matrices. Since we do not want to restrict ourselves to small systems we need an approximate scheme which can deal efficiently with huge systems. The Dyson equation can be solved in a perturbative fashion by realizing that it can be expressed in the open form
 \begin{align}
    \label{eq: grn dyson open}
    \mathbf{G} = \mathbf{g} + \mathbf{g} \mathbf{t} \mathbf{g} + \mathbf{g} \mathbf{t} \mathbf{g} \mathbf{t} \mathbf{g} + \ldots
    \; .
\end{align}
This open form serves as the foundation for the perturbative approach, enabling an analytical approximation of the full Green function by neglecting higher-order terms in $\mathbf{t}$.

Starting from eq. \ref{eq: grn dyson open}), we consider only the first non-vanishing correction to the density of states (dos). The density of states is related to Green function:
 \begin{align}
    \label{eq: grn dos}
    dos_{j}(\epsilon) = \frac{-1}{\pi}\Im\, \text{Tr}_{\sigma}(\mathbf{G}_{jj}(\epsilon))\;.
\end{align}
$\Im$ stands for the imaginary part, $\text{Tr}_{\sigma}$ represents the trace over the spin-channel and $j$ is labelling the sites. By combining the definitions of the dos (eq. \ref{eq: grn dos}) and the Dyson equation (eq. \ref{eq: grn dyson open}), we arrive at a perturbative expansion for the dos at site $j$
\begin{align}
\label{eq: grn dps approx}
   dos_{j}(\epsilon)  &\approx
    \frac{- 1}{\pi}\Im\,  \text{Tr}_{\sigma}( \mathbf{g}_{j}+\mathbf{g}_{j} \, \mathbf{t}_{j,j} \, \mathbf{g}_{j} + \sum_{i} \mathbf{g}_{j} \, \mathbf{t}_{j,i} \, \mathbf{g}_{i} \, \mathbf{t}_{i,j} \, \mathbf{g}_{j} \,+ O(\mathbf{t}^3)) \\
     &\approx      \frac{- 1}{\pi}\Im\,  \text{Tr}_{\sigma}( \mathbf{g}_{j}+ \sum_{i} \mathbf{g}_{j} \, \mathbf{t}_{j,i} \, \mathbf{g}_{i} \, \mathbf{t}_{i,j} \, \mathbf{g}_{j} \,+ O(\mathbf{t}^3)) \\
     \label{eq: grn expansion charge}
    & \approx dos_{j,0}(\epsilon) 
    -\frac{ 1}{\pi}\sum_{i} \Im \, \text{Tr}_{\sigma}(  \mathbf{g}_{j} \, \mathbf{t}_{j,i} \, \mathbf{g}_{i} \, \mathbf{t}_{i,j} \, \mathbf{g}_{j} \,)+ O(\mathbf{t}^3)\;.
\end{align}
Here, we utilize the fact that hopping is purely non-local, meaning that $\mathbf{t}_{i,i} = 0$. As a result, the first-order correction in $t$ vanishes, allowing us to focus on the second-order correction. Naturally, the accuracy of the results would depend on the amplitude of the hopping. It is thus important to exercise caution when interpreting the results, as even non-physical phenomena may arise when a strong hopping $\mathbf{t}$ is employed. However, for small values of $\mathbf{t}$, this approximation can provide reasonably accurate quantitative outcomes for the utilized Hamiltonian. \newline
Let us examine the term in detail: $\text{Tr}_{\sigma}( \mathbf{g}_{j} , \mathbf{t}_{j,i} , \mathbf{g}_{i} , \mathbf{t}_{i,j} , \mathbf{g}_{j} ) = \xi$. We will explicitly derive all contribution to $\xi$ in a general fashion. The non-interacting Green function $g$ can always be written in the form
\begin{equation}
\label{eq: grn1}
    \mathbf{g}_{j}(\epsilon)= a(\epsilon) + b(\epsilon) (\mathbf{e}_{j} \cdot \mathbf{\sigma} )
\end{equation}
 and similarly for the hopping
 \begin{equation}
    \mathbf{t}_{ij}= c + d (\mathbf{n}_{ij} \cdot \mathbf{\sigma} )
\end{equation}
with $\mathbf{n}_{ij} = - \mathbf{n}_{ji}$. We drop in the notation the energy $\epsilon$ and can expand $\xi$
 \begin{equation}
    \xi= \text{Tr}_{\sigma}\big((a + b\, \mathbf{e}_{j} \cdot \mathbf{\sigma} )(c + d\, \mathbf{n}_{ij} \cdot \mathbf{\sigma} )(a + b\, \mathbf{e}_{i} \cdot \mathbf{\sigma} )(c + d\, \mathbf{n}_{ji} \cdot \mathbf{\sigma} )(a + b\, \mathbf{e}_{j} \cdot \mathbf{\sigma} )\big)\;.
\end{equation}

By making use of $(\mathbf{e}_{i} \cdot \sigma)(\mathbf{e}_{j}\cdot \sigma) = \mathbf{e}_{i} \cdot \mathbf{e}_{j} + i (\mathbf{e}_{i} \times \mathbf{e}_{j} )\cdot \sigma  $ and broken inversion symmetry $\mathbf{n}_{ij}=-\mathbf{n}_{ji}$ we expand this product of Pauli matrices and sort the results by the number of $b$'s and $d$'s:
\begin{table}[h!]
\centering
\begin{tabular}{ c }
 \begin{tikzpicture}
    \matrix (contribution) [matrix of nodes,nodes={minimum width=1cm,minimum height=1cm},inner sep=2mm]
    {
    & \node{0}; & \node{1};      & \node{2};       & \node{3};  \\
  \node {0}; & $2 a^3 c^2$ & 0 & $2 a b^2 c^2(1+2 \mathbf{e}_{j} \cdot \mathbf{e}_{i})$ & 0 \\
  \node {1}; & 0                             & 0 & $i 8 a b^2 c d (\mathbf{e}_{i} \times \mathbf{e}_{j})\cdot \mathbf{n}_{ji}$ & 0 \\
  \node {2}; & $-2 a^3 d^2$ & 0 & $-2 a b^2 d^2 (1+ 4 (\mathbf{e}_{j}\cdot \mathbf{n}_{ji}) ( \mathbf{e}_{i} \cdot \mathbf{n}_{ji}) -2 \mathbf{e}_{j} \cdot \mathbf{e}_{i})$ & 0 \\
  };
    
    \node[anchor=south, xshift = 5mm, yshift = -5 mm, rotate=90] at (contribution.west) {Order of $d$ (SOC) };
    \node[anchor=south, xshift = 6mm, yshift = -6 mm] at (contribution.north) {Order of $b$ (magnetic Green function)};
\end{tikzpicture}  \\
\end{tabular}
\caption{Tabulation showcasing the various contributions to $\xi$ upon incorporating the hopping and Hamiltonian for the sites. The rows are arranged based on the count of $d$'s which stands for SOC-related hopping events in our framework, while the columns are organized by the number of $b$'s, i. e. the amount of magnetic scattering occurrences at a site.}
\label{tab: grn expansion}
\end{table}

Important for this work is the column with $b=2$ in \ref{tab: grn expansion}. There one can identify two isotropic ($\mathbf{e}_{j} \cdot \mathbf{e}_{i}$), a chiral ($(\mathbf{e}_{i} \times \mathbf{e}_{j})\cdot \mathbf{n}_{ji}$) and a anisotropic ($(\mathbf{e}_{j}\cdot \mathbf{n}_{ji}) ( \mathbf{e}_{i} \cdot \mathbf{n}_{ji})$) contribution to $\xi$. Till now, we only looked at the dos, we can easily extract the charge by performing a energy integration of the dos
\begin{align}
   \rho_{j}  &=
    \int_{-\infty}^{\epsilon_{F}}\,dos_{j}(\epsilon) \text{d}\epsilon  \\
    & \approx \rho_{j,0}
    -\frac{ 1}{\pi}\sum_{i} \Im \, \int_{-\infty}^{\epsilon_{F}}\, \text{Tr}_{\sigma}(  \mathbf{g}_{j} \, \mathbf{t}_{j,i} \, \mathbf{g}_{i} \, \mathbf{t}_{i,j} \, \mathbf{g}_{j} )\text{d}\epsilon+ O(\mathbf{t}^3)\;.
\end{align}
We can easily perform the energy integration analytically, since in \ref{tab: grn expansion} only $a$ and $b$ depend on the energy, the integration simplifies significantly to $\int_{-\infty}^{\epsilon_{F}}\, a b^2  \text{d}\epsilon$ for the magnetic charge corrections. Comparing the definition of $a$ and $b$ in eq. \ref{eq: grn1} with the Hamiltonian, we obtain 
\begin{align}
    a &= \frac{ g_{\uparrow\uparrow,loc} +  g_{\downarrow\downarrow,loc}}{2} = 
    \frac{1}{2}\frac{1}{\epsilon-(\epsilon_{d}- i \Gamma - U m)}+\frac{1}{2}\frac{1}{\epsilon-(\epsilon_{d}- i \Gamma + U m)} \\
    b &= \frac{ g_{\uparrow\uparrow,loc} -  g_{\downarrow\downarrow,loc}}{2} = 
    \frac{1}{2}\frac{1}{\epsilon-(\epsilon_{d}- i \Gamma - U m)}-\frac{1}{2}\frac{1}{\epsilon-(\epsilon_{d}- i \Gamma + U m)}\,.
\end{align}

By inserting these definitions we can perform the integration
\begin{align}
\label{eq: grn integrating greenfunction}
    \int_{-\infty}^{\epsilon_{F}}ab^2\text{d}\epsilon &= \frac{1}{8}\int_{-\infty}^{\epsilon_{F}}( g_{\uparrow\uparrow,loc}(\epsilon) +  g_{\downarrow\downarrow,loc}(\epsilon)) ( g_{\uparrow\uparrow,loc}(\epsilon) -  g_{\downarrow\downarrow,loc}(\epsilon))^2\text{d}\epsilon \\
    &=\frac{1}{16}(2\, g_{ \uparrow\uparrow,loc}(\epsilon_{F}) g_{ \downarrow\downarrow,loc}(\epsilon_{F})-g_{\uparrow\uparrow,loc}^{2}(\epsilon_{F}) - g_{\downarrow\downarrow,loc}^{2}(\epsilon_{F}))\;.
\end{align}

Thereby we have used the following helpful relations
\begin{align}
    \frac{g_{\uparrow\uparrow,loc}-g_{\downarrow\downarrow,loc}}{2 U  m} =  g_{\uparrow\uparrow,loc} g_{\downarrow\downarrow,loc} \\
    \partial_{\epsilon} g_{ss,loc} = - (g_{ss,loc})^{2}, s = \downarrow, \uparrow\;.
\end{align}

In order to verify our model, we test our approximation on a dimer, for which we can calculate the Green function $G$ analytically. We re-scale the hopping strength $ t_{dimer} = \sqrt{6} t \approx -0.17$ eV to account for the reduced number of nearest neighbours exactly for 2nd-order contribution in $t_{dimer}$. The strength of higher order contribution in $t_{dimer}$ are increased, meaning that if there are negligible in the dimer case than our approximation are also valid in the 3 dimensional case with 6 nearest neighbours and lower hopping strength $t$.

\begin{figure}[H]
    \centering
    \includegraphics[width=.99\textwidth]{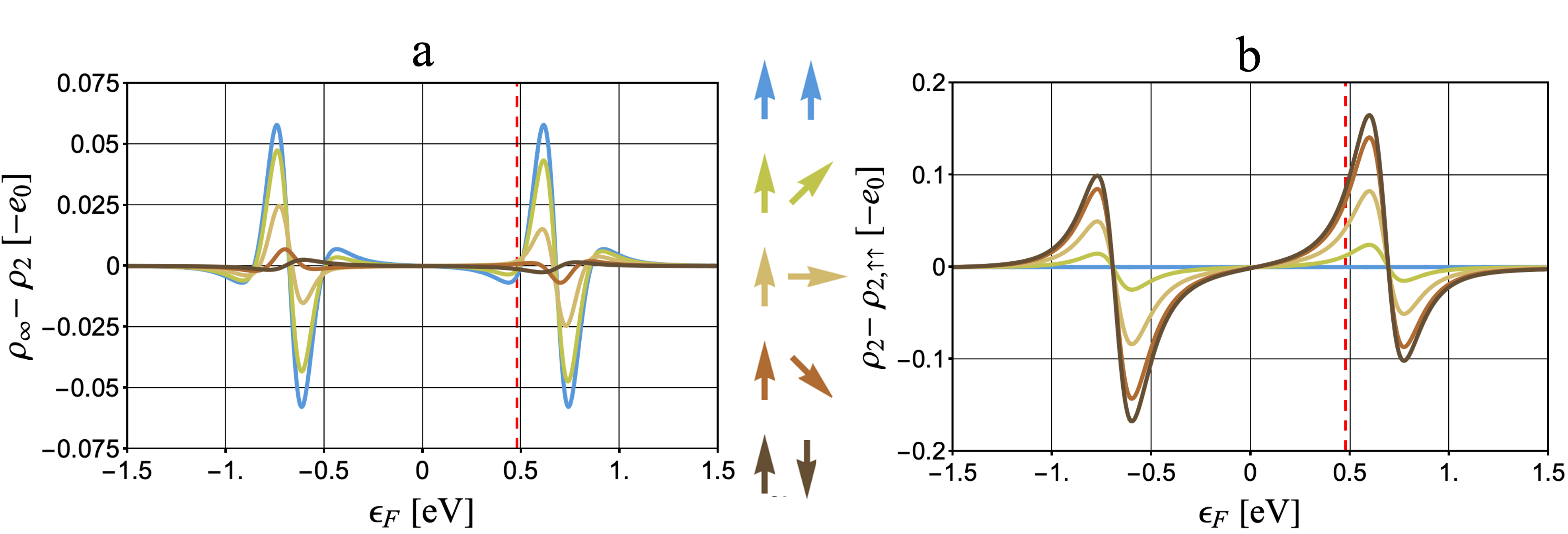}
    \caption{\textbf{Exact versus perturbative scheme to evaluate the charge of misaligned magnetic moments of a dimer.} Error of the perturbative method (a) and amplitude of the non-collinearity-induced charge (b) as function of the Fermi energy for different magnetic alignments of the dimer. In (a) the charge difference obtained from the two methods: solving fully the Dyson equation leading to $\rho_\infty$ is compared to that based on the perturbative scheme $\rho_2$. In (b) the non-collinearity-induced charge is evaluated from the perturbative scheme. The dashed line indicates an example of the Fermi energy hosting a large non-collinearity-induced charge while minimizing the error of the perturbative method.}
    \label{fig: sub 1}
\end{figure}

In Fig.~\ref{fig: sub 1} a, we compare the charge density $\rho_\infty$ obtained by solving the full Dyson equation to that obtained from perturbation theory up to second order in the hopping $\rho_2$. Clearly the changes are minimal, with the largest errors occurring around the energies associated to the electronic orbitals. Indeed, shifts induced at the level of resonances can be large and thus more difficult to grasp using perturbative schemes. Fig.~\ref{fig: sub 1} b shows the non-collinearity-induced charge, evaluated up to second order in the hopping. This helps to visualize the range of Fermi energies where the largest magnitude of the induced charge are localized. Clearly the signal is enhanced when close to the electronic orbitals energies, where errors from the schemes emerge.
However, it is possible to find Fermi energies that have an acceptable error and still provide a strong contrast through rotation, an example of this is the red dashed line in \ref{fig: sub 1}. 

\section*{Supplementary Note 3 - Micromagnetics.}
\label{supplementary_micromagnetics}

In this part we re-express the atomistic isotropic, chiral and anisotropic contributions in the micromagnetic limit~\cite{article_Brown1963} terms of a smooth varying magnetization $\mathbf{m}(\mathbf{r})$. In the discrete lattice model we denoted with site $j$ the nearest neighbours of site $i$. Using the continuous magnetization we can express the magnetization at site $j$ as a Taylor-expansion around site $i$ (for simplicity we assumed site $i$ is located at $\mathbf{r}$ and site $j$ is located at $\mathbf{r} + a \mathbf{e}_{x}$

\begin{align}
    \label{eq: grn micro1}
    &\mathbf{e}_{i} \rightarrow \mathbf{m}(\mathbf{r}), \\
    & \mathbf{e}_{j} \rightarrow \mathbf{m}(\mathbf{r}) + a \partial_{x} \mathbf{m}(\mathbf{r}) + \frac{a^{2}}{2} \partial_{x}^{2} \mathbf{m}(\mathbf{r})
    + ... \, .
\end{align}

Thereby stands $a$ for the lattice constant. With these expansion we translate the dependency on the misalignment for a simple cubic lattice into the micromagnetic limit 

\begin{align}
       isotropic(x,z,y) \propto  \sum_{j} (\mathbf{e}_{i}\cdot \mathbf{e}_{j}) &\rightarrow 6 + a^{2}\, \mathbf{m}^{\text{T}} \left[\begin{array}{rrr} 
            \partial_{x}^{2} & 0 & 0 \\ 
            0 & \partial_{y}^{2} & 0 \\ 
            0 & 0 & \partial_{z}^{2} \\ 
        \end{array}\right]
        \mathbf{m}\\
        chiral(x,z,y) \propto  \sum_{j} (\mathbf{e}_{i}\times \mathbf{e}_{j})\cdot\mathbf{n}_{ji} &\rightarrow 2 a\, \mathbf{m}^{\text{T}} \left[\begin{array}{rrr} 
            0 & 0 & -\partial_{x} \\ 
            0 & 0 & -\partial_{y} \\ 
            \partial_{x} & \partial_{y} & 0 \\ 
        \end{array}\right]
        \mathbf{m}\\
            anisotropic(x,z,y) \propto  \sum_{j} (\mathbf{e}_{j}\cdot\mathbf{n}_{ji})(\mathbf{e}_{i}\cdot\mathbf{n}_{ji}) &\rightarrow 2 (1-m_{z}^{2})+ a^{2}\,\mathbf{m}^{\text{T}} \left[\begin{array}{rrr} 
            \partial_{y}^{2} & 0 & 0 \\ 
            0 & \partial_{x}^{2} & 0 \\ 
            0 & 0 & 0 \\ 
        \end{array}\right]
        \mathbf{m} \, .
    \label{eq: grn microcontribution}
\end{align}

Naturally, this micromagnetic picture is applicable for large spin textures and we restrict ourselves to the first nontrivial term of the Taylor-expansion that contains derivatives. The isotropic contribution contains second-order derivatives of the magnetization and the isotropic nature can be seen at the symmetric structure. The chiral contribution does not posses this symmetric structure and contains first-order derivatives. Particularly interesting is the anisotropic contribution which contains the term $(1-m_{z}^{2})$. This term does not contain a derivative, i. e. it is not sensitive to the misalignment of the magnetization and has the shape of a anisotropy energy. However, there is a correction with second-order derivatives which are sensitive to the misalignment of the magnetization. We expect for large spin textures, where the second-order derivatives are negligible, only a local anisotropic behavior from the anisotropic contribution.

\section*{Supplementary Note 4 - Electronic contribution to electron holography.}
The electronic contribution demands the development of a new method, one that is consistent with the approach used for the magnetic part. 
We recall the definition of the electronic contribution to the phase-image and insert the Poisson’s equation:
\begin{align}
    \varphi_{e}(x,y) &= C_{E} \int \Phi(x,y,z) \text{d}z = \frac{C_{E}}{ 4 \pi \epsilon_{0} \epsilon_{r}}  \int \int \frac{\rho(\mathbf{r}')}{\lvert\mathbf{r}-\mathbf{r}'\rvert}\text{d}^3r' \, \text{d}z ,
    \label{eq: phase shift divergence}
\end{align}
which involves a fundamental problem:  the long-range nature inherent to the Coulomb potential causes divergence of the phase-shift. Therefore, we incorporate Thomas–Fermi screening with the screening parameter $k_{tf}$

\begin{align}
  \varphi_{e}(x,y) &= \frac{C_{E}}{ 4 \pi \epsilon_{0} \epsilon_{r}} \int \rho(\mathbf{r}') \int \frac{e^{-\lvert\mathbf{r}-\mathbf{r}'\rvert k_{tf}}}{\lvert\mathbf{r}-\mathbf{r}'\rvert} \text{d}z \, \text{d}^3r' .
  \label{eq: yukawa}
\end{align}

There is still a critical issue to address. Upon careful examination of eq. \ref{eq: phase shift divergence} or  eq. \ref{eq: yukawa}  it becomes apparent that the integral diverges for $\mathbf{r} = \mathbf{r}'$. This divergence stems from the singular nature of the point charge. To resolve this challenge, we address the point charge singularity by spreading it into a uniform charge distribution within a sphere of radius $R$

\begin{align}
  \varphi_{e}(x,y) &= \frac{C_{E}}{ 4 \pi \epsilon_{0} \epsilon_{r} V_{sphere}(R)} \int \rho(\mathbf{r}') \int \int \theta(R - \lvert \Tilde{\mathbf{r}} \rvert ) \frac{e^{-k_{tf}\lvert\mathbf{r}-\mathbf{r'}-\Tilde{\mathbf{r}}\rvert}}{\lvert\mathbf{r}-\mathbf{r'}-\Tilde{\mathbf{r}}\rvert} \text{d}^{3}\Tilde{r}\, \text{d}z \, \text{d}^3r' .
  \label{eq: yukawa2}
\end{align}

Here, $\theta$ denotes the Heaviside function and $V_{sphere}(R)$ corresponds to the volume of a sphere with radius $R$. The integration with respect to $\Tilde{\mathbf{r}}$ can be treated analytical, while the remaining integration are solved numerically.

\section*{Supplementary References}
\bibliography{references}